\documentclass{ws-ijmpd}
\usepackage[super,compress]{cite}
\usepackage{graphicx}
\usepackage{dcolumn}
\usepackage{bm}
\usepackage{amssymb,amsmath
	,accents}
\allowdisplaybreaks[4]
\usepackage{amscd,amsfonts}
\usepackage{mathrsfs}
\usepackage{epsfig}
\usepackage{epstopdf}
\DeclareGraphicsExtensions{.eps,.pdf,.png,.jpg,.jpeg}
\begin{document}
\markboth{Muhammad Usman and Asghar Qadir}
{Dark energy from non-degenerate Higgs-vacuum}
%
\catchline{}{}{}{}{}
%
\title{Dark energy from non-degenerate Higgs-vacuum}
\author{Muhammad Usman}
\address{Department of Physics, \\COMSATS University, Islamabad 44000, Pakistan\\
muhammad\_usman\_sharif@yahoo.com}
\author{Asghar Qadir}
\address{Department of Physics, \\ School of Natural Sciences $\left(\text{SNS}\right)$,\\ National University of Sciences and Technology $\left(\text{NUST}\right)$,\\ Sector H-12, Islamabad 44000, Pakistan \\
asgharqadir46@gmail.com}
\maketitle
\begin{history}
\received{12 June 2019}
\accepted{13 June 2019}
\end{history}
\begin{abstract}
Scalar fields are favorite among the possible candidates for the dark energy. Most frequently discussed are those with degenerate minima at $\pm \phi_{min}$. In this paper, a slightly modified two-Higgs doublet model 
is taken to contain the Higgs field(s) as the dark energy candidate(s).
The model considered has two non-degenerate minima at $\phi_{\pm}$, instead of one degenerate minimum at $\pm\phi_{min}$. The component fields of one SU(2) doublet ($\phi_1$) act as the standard model (SM) Higgs, while the component fields of the second doublet ($\phi_2$) are taken to be the dark energy candidates (lying in the true vacuum). It is found that one \emph{can} arrange for late time acceleration (dark energy) by using an SU(2) Higgs doublet, whose vacuum expectation value is zero, in the quintessential regime.
\end{abstract}
\ccode{PACS numbers: 95.36.+x; 98.80.-k; 12.60.Fr; 14.80.Cp}
\section{Introduction}
Scalar fields are among the favourite candidates for dark energy and were first used for inflation by Alan Guth \cite{guth} and Andre Linde \cite{Linde,LINDE1982389}. In our context they are taken to be new fields (with no connection to Particle Physics whatsoever). We feel that the tried and tested Physics (SM) may be able to provide the dark energy.

The dynamics of the current Universe is best described by the Friedmann equations which are
\begin{eqnarray}
	H^2 &=& \dfrac{8\pi G}{3}\rho-\dfrac{\kappa}{a^2}~, \label{1stFriedmannequation}
	\\
	\dfrac{\ddot{a}}{a} &=& -\dfrac{1}{6}\left(1+3\omega_{eff}\right)\rho~, \label{2ndFriemannequation}
\end{eqnarray}
where $a$ is the scale factor, $H=\dot{a}/a$ is the Hubble parameter, $\rho$ is the total energy density of the Universe and $\kappa$ is the spatial curvature. In deriving the above equations, the barotropic equation of state $P=\omega\rho$ has been used; $P$ being the pressure and $\omega$ the equation of state parameter.  Here, and throughout, Planck units $\hbar=c=1$ and $(8\pi G)^{-1/2}=M_P$ have been used and $M_P$ has been taken to be 1. From eq. (\ref{2ndFriemannequation}), we see that $\ddot{a}>0$ (accelerated expansion of the Universe) when $\omega_{eff}<-\frac{1}{3}$.
Scalar fields are categorized as:
(1) \emph{Quintessence fields} ($-1<\omega<-1/3$);
(2) \emph{Phantom fields} ($\omega<-1$). 
When $\omega=-1$ the energy density remains constant as the Universe expands. 

In this paper, we assume that the present state of the Universe is described by an inert uplifted double well two-Higgs doublet model in which both doublets lie in their true vacuum, and the vacuum expectation value (VeV) of the second doublet is zero. 
The results obtained favor the accelerated expansion of the Universe in the quintessence regime. 
The paper is organized as follows: in section (\ref{sec:2}) we describe the uplifted double well two-Higgs doublet model (UDW-2HDM), in section (\ref{sec:3}) we perform the analysis of UDW-2HDM Higgs field(s) as dark energy candidate(s). In section (\ref{sec:4}), we give the decay(s) of Higgs field(s). Section (\ref{sec:Conclusion}) is the conclusion.
In contrast to this article, we previously have demonstrated that an ever accelerated expansion of the Universe can be obtained by the usual two Higgs doublet model (2HDM) of Particle Physics if the second phase transition has not occurred in the 2HDM \cite{doi:10.1142/S0218271817410036}.
\section[UDW type 2HDM]{Uplifted double well two-Higgs double model}\label{sec:2}
\begin{figure}[h!]
	\centering
	\includegraphics[scale=.35]{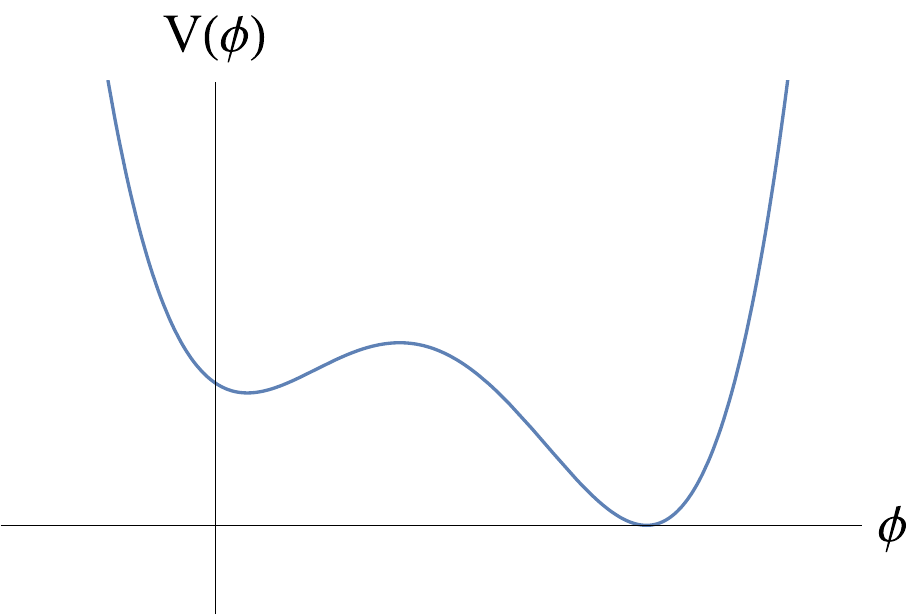}
	\caption{The uplifted double well potential}
	\label{fig:potential}
\end{figure}

The Higgs field Lagrangian is
\begin{equation}\label{LH}
	\mathcal{L}_{Higgs}=T_{H}-V_{H}~.
\end{equation}
The kinetic and potential terms in UDW-2HDM are
\begin{equation}\begin{array}{rcl}\label{TH}
		T_H= ({D_1}_{\mu}\phi_1)^\dagger ({D_1}^{\mu}\phi_1) &+& ({D_2}_{\mu}\phi_2)^\dagger ({D_2}^{\mu}\phi_2) \\ & & +\left [ {\chi}({D_1}_{\mu}{\phi}_{1})^{\dagger}({D_2}^{\mu}{\phi}_{2})+{\chi^{*}}({D_2}_{\mu}{\phi}_{2})^{\dagger}({D_1}^{\mu}{\phi}_{1}) \right ],
	\end{array}
\end{equation}
and
\begin{equation}\begin{array}{rcl}\label{VH}
		V_H & = &
		\rho_1\exp(\Lambda_1m_{11}^2(\phi_1-\phi_{1_0})^\dagger(\phi_1-\phi_{1_0}))+ \rho_3\exp\left(\dfrac{1}{2}\Lambda_3\lambda_1((\phi_1-\phi_{1_0})^\dagger(\phi_1-\phi_{1_0}))^2\right) 
		\\ & & \hspace{-0.25cm}
		+\rho_2 \exp(\Lambda_2 m_{22}^2(\phi_2-\phi_{2_0})^\dagger(\phi_2-\phi_{2_0}))+\rho_4\exp\left(\dfrac{1}{2}\Lambda_4\lambda_2((\phi_2-\phi_{2_0})^\dagger(\phi_2-\phi_{2_0}))^2\right) \\ & & \hspace{-0.25cm}
		+\lambda_3(\phi_1^\dagger\phi_1)(\phi_2^\dagger\phi_2)+\lambda_4(\phi_1^\dagger\phi_2)(\phi_2^\dagger\phi_1)+\left[{m_{12}^2}(\phi_1^\dagger \phi_2) \right. +\dfrac{\lambda_5}{2}(\phi_1^\dagger\phi_2)^2 
		\\ & & \hspace{6.05cm}
		\left. +\lambda_6(\phi_1^\dagger\phi_1)(\phi_1^\dagger\phi_2)+ \lambda_7(\phi_2^\dagger\phi_2)(\phi_1^\dagger\phi_2)+\text{h.c.}\right],
	\end{array}
\end{equation}
where
\begin{equation*}
	{D_1}_\mu=\partial_\mu+\dot{\iota}\dfrac{g_1}{2}\sigma_i {W^i}_\mu+\dot{\iota}\dfrac{g_1'}{2}B_\mu,
\end{equation*}
\begin{equation*}
	{D_2}_\mu=\partial_\mu+\dot{\iota}\dfrac{g_2}{2}\sigma_i {W^i}_\mu+\dot{\iota}\dfrac{g_2'}{2}B_\mu,
\end{equation*}
\begin{equation*}
	\phi_{i}=
	\begin{bmatrix}
		\phi^{+}_{i} \\
		\eta_i + \dot{\iota}\chi_i +\nu_i \\
	\end{bmatrix}
	\text{ , \qquad}
	\phi_{i_0}=
	\begin{bmatrix}
		0 \\
		\tau_i \\
	\end{bmatrix}	
	\text{\qquad and \qquad}
	\phi_{i}^{\dagger}=
	\begin{bmatrix}
		\phi^{-}_{i} & \eta_i - \dot{\iota}\chi_i +\nu_i
	\end{bmatrix}.
\end{equation*}
The dimensions of the different quantities are
$[\rho_i]^{-1}=[\Lambda_i]=[L]^4,\text{\space} [m_{ii}^2]=[L]^{-2},\text{\space} [\phi_i]=[L]^{-1}\text{\space and \space}[\lambda_i]=[L]^0,$
where ``$L$'' denotes length. The Higgs fields $\phi^{+}_{i}$, $\phi^{-}_{i}$, $\eta_i$ and $\chi_i$ are hermitian ($\phi^{\pm}_{i}$ are charged whereas the others are neutral), $\nu_i$ is the VeV of the doublet $\phi_i$, $\phi_{i_0}$ in the potential is the true minimum of the field $\phi_i$. Here we also assume that both the Higgs doublets are coupled with the gauge fields differently. In this way, we can suppress the interaction of dark energy Higgs with the gauge bosons. 
The shape of the potential in this model is shown in fig. (\ref{fig:potential}). 

Since we want the Higgs field (lying in the false vacuum) to live for not less than the current age of the Universe, a stable Higgs field is required. This can be achieved by imposing a discrete $Z_2$ symmetry.

There are two types of $Z_2$ symmetry breaking: $\left.1\right)$ soft; and $\left.2\right)$ hard. The term containing $m_{12}^{2}$ describes the soft $Z_2$ symmetry breaking, whereas the terms containing $\lambda_{6}$ and $\lambda_{7}$ describe the hard $Z_2$ symmetry breaking \footnote{When $Z_2$ symmetry is broken by $(\phi_i^\dagger\phi_j)$ type terms then it is said to be softly broken and when $Z_2$ symmetry is broken by $(\phi_i^\dagger\phi_j)(\phi_k^\dagger\phi_l)$ type terms then $Z_2$ is said to be hardly broken}. In the absence of these terms along with no cross kinetic term i.e. $\chi=0$, the UDW-2HDM's Higgs Lagrangian has a perfect $Z_{2}$ symmetry \cite{Ginzburg-2}. There are two $Z_2$ symmetries corresponding to the UDW-2HDM doublets:
\begin{eqnarray}
	\text{I:\qquad\qquad}\phi_1 \longrightarrow &-\phi_1~, \text{\qquad\qquad} \phi_2 \longrightarrow &\text{\space\space}\phi_2~, \label{1stZ2symmetry}
	\\
	\text{ \quad\qquad\qquad}\phi_{1_0} \longrightarrow &-\phi_{1_0}~, \text{\qquad\qquad} \phi_{2_0} \longrightarrow &\text{\space\space}\phi_{2_0}~; \nonumber		 
	\\
	\text{II:\qquad\qquad}\phi_1 \longrightarrow &\text{\space\space}\phi_1~, \text{\qquad\qquad} \phi_2 \longrightarrow &-\phi_2~,	\label{2ndZ2symmetry}
	\\
	\text{ \qquad\qquad}\phi_{1_0} \longrightarrow &\text{\space\space}\phi_{1_0}~, \text{\qquad\qquad} \phi_{2_0} \longrightarrow &-\phi_{2_0}~.	\nonumber
\end{eqnarray}
For $Z_{2}$ symmetry see \cite{Ginzburg-2}.
\subsection{Minimizing the Higgs potential}
The extrema of the potential are found by taking
\begin{equation}\label{extremaconditions}
	\dfrac{\partial V_H}{\partial \phi_1} \Big{|}_{\phi_1=\left\langle \phi_1\right\rangle }=\dfrac{\partial V_H}{\partial \phi_1^\dagger}\Big |_{\phi_1^\dagger= \left\langle \phi_1^\dagger\right\rangle}=0 \text{\qquad and \qquad} \dfrac{\partial V_H}{\partial \phi_2}\Big |_{\phi_2=\left\langle \phi_2\right\rangle }=\dfrac{\partial V_H}{\partial \phi_2^\dagger}\Big |_{\phi_2^\dagger=\left\langle \phi_2^\dagger\right\rangle }=0.
\end{equation}

The most general solution of the conditions (\ref{extremaconditions}) is
\begin{equation*}\label{VeV}
	\left\langle \phi_1\right\rangle =\frac{1}{\sqrt{2}}\begin{pmatrix}
		0 \\ \nu_1
	\end{pmatrix}
	\text{\qquad and \qquad}
	\left\langle \phi_2\right\rangle =\frac{1}{\sqrt{2}}\begin{pmatrix}
		u \\ \nu_2
	\end{pmatrix}.
\end{equation*}
The first solution of extrema has been taken to be similar to the Higgs vacuum in the SM and the second one is the most general that could occur. One needs to keep in mind that now $\nu^2=\nu_1^2+\left| \nu_2^2\right|+u^2$, where $\nu=1 / \sqrt[4]{2{G_F}^2}\approx246\text{GeV}$ is the VeV of the Higgs field in the SM. 

If $u\neq 0$ the non-zero value of $u$ will contribute to the ``charged'' type dark energy, which has not been observed. To avoid this, we choose $u=0$. From the extrema conditions given by eq. (\ref{extremaconditions}), we can determine the values of $\nu_1$ and $\nu_2$ \cite{2hdmc,Eriksson2010985,Ginzburg-1}, solving eq. (\ref{extremaconditions}) for the potential given by eq.(\ref{VH}) leads to
\begin{equation}\begin{array}{rcl}\label{1stexpcondition}
		& & \hspace{-1cm}\rho_1\Lambda_1m_{11}^2(\nu_1-\tau_1)\exp\left( \Lambda_1m_{11}^2(\nu_1-\tau_1)^2 \right) \\ & &
		\hspace{-1cm}+ \rho_3\Lambda_3\lambda_1\nu_1(\nu_1^2-\tau_1^2)\exp\left( \dfrac{\Lambda_3\lambda_1}{2}(\nu_1^2-\tau_1^2)^2 \right) \\ & & \hspace{-1cm}+{{m_{12}^2}}^{*}\nu_2+(\lambda_3+\lambda_4+\lambda_5^{*})\nu_1\nu_2^2+(\lambda_6+2\lambda_6^{*})\nu_1^2\nu_2+\lambda_7^{*}\nu_2^3=0,
	\end{array}
\end{equation}
\begin{equation}\begin{array}{rcl}\label{2ndexpcondition}
		& & \hspace{-1cm}\rho_2\Lambda_2m_{22}^2(\nu_2-\tau_2)\exp\left( \Lambda_2m_{22}^2(\nu_2-\tau_2)^2 \right) \\ & &
		\hspace{-1cm}+ \rho_4\Lambda_4\lambda_2\nu_2(\nu_2^2-\tau_2^2)\exp\left( \dfrac{\Lambda_4\lambda_2}{2}(\nu_2^2-\tau_2^2)^2 \right) \\ & & \hspace{-1cm}+{{m_{12}^2}}\nu_1+(\lambda_3+\lambda_4+\lambda_5)\nu_1^2\nu_2+\lambda_6\nu_1^3+(2\lambda_7+\lambda_7^{*})\nu_1\nu_2^2=0.
	\end{array}
\end{equation}

The solution of the above equations is practically impossible. Thus, truncating the terms to fourth order in the exact $Z_2$ symmetric potential
, which makes the lightest Higgs field stable and a dark energy candidate before phase transition. 
Now applying the minimization conditions on the new potential, 
%
alongside imposing $\lambda_3+\lambda_4+\lambda_5=0$ \cite{USMAN2015arXiv150607099U}$^{,}$\footnote{With $\lambda_3+\lambda_4+\lambda_5\neq0$ the solution to the VeV contain complex part} we get four solutions for $\nu_1$ and $\nu_2$, which are:
\begin{eqnarray}
	&\nu_1 = \tau_1
	~, \text{ } &\nu_2 = \tau_2 ~; \label{electroweaksymmetricvacuum}
	\\
	&\nu_1 = \tau_1 ~, \text{ } &\nu_2 =\dfrac{2 m_{22}^4 \Lambda _2^2 \rho _2 \tau _2-\lambda _2 \Lambda _4 \rho _4 \tau _2-\sqrt{\Xi_2}}{2 \left(m_{22}^4 \Lambda _2^2 \rho _2+\lambda _2 \Lambda _4 \rho _4\right)} ~; \label{inertvacuum1}
	\\
	&\nu_1 =\dfrac{2 m_{11}^4 \Lambda _1^2 \rho _1 \tau _1-\lambda _1 \Lambda _3 \rho _3 \tau _1-\sqrt{\Xi_1}}{2 \left(m_{11}^4 \Lambda _1^2 \rho _1+\lambda _1 \Lambda _3 \rho _3\right)} ~, \text{\qquad} &\nu_2 = \tau_2 ~; \label{inertvacuum2}
	\\
	&\nu_1 =\dfrac{2 m_{11}^4 \Lambda _1^2 \rho _1 \tau _1-\lambda _1 \Lambda _3 \rho _3 \tau _1-\sqrt{\Xi_1}}{2 \left(m_{11}^4 \Lambda _1^2 \rho _1+\lambda _1 \Lambda _3 \rho _3\right)} ~, \text{\qquad} &\nu_2 =\dfrac{2 m_{22}^4 \Lambda _2^2 \rho _2 \tau _2-\lambda _2 \Lambda _4 \rho _4 \tau _2-\sqrt{\Xi_2}}{2 \left(m_{22}^4 \Lambda _2^2 \rho _2+\lambda _2 \Lambda _4 \rho _4\right)} ~; \label{mixedvacuum}
\end{eqnarray}
where
$$\Xi_1=-4 m_{11}^6 \Lambda _1^3 \rho _1^2-4 m_{11}^2
\lambda _1 \Lambda _1 \Lambda _3 \rho _1 \rho _3-8 m_{11}^4 \lambda _1 \Lambda _1^2 \Lambda _3 \rho _1 \rho _3 \tau _1^2+\lambda _1^2 \Lambda _3^2
\rho _3^2 \tau _1^2,$$
and
$$\Xi_2=-4 m_{22}^6 \Lambda
_2^3 \rho _2^2-4 m_{22}^2 \lambda _2 \Lambda _2 \Lambda _4 \rho _2 \rho _4-8 m_{22}^4 \lambda _2 \Lambda _2^2 \Lambda _4 \rho _2 \rho _4 \tau _2^2+\lambda
_2^2 \Lambda _4^2 \rho _4^2 \tau _2^2.$$

The Higgs doublets, when the vacuum is given by eq. (\ref{electroweaksymmetricvacuum}), are
\begin{equation}
	\phi_{1}=
	\begin{bmatrix}
		\phi^{+}_{1} \\
		\eta_1 + \dot{\iota}\chi_1 + \tau_1 \\
	\end{bmatrix},
	\text{\qquad\qquad}
	\phi_{2}=
	\begin{bmatrix}
		\phi^{+}_{2} \\
		\eta_2 + \dot{\iota}\chi_2 + \tau_2\\
	\end{bmatrix}.
\end{equation}
If we choose $\tau_2=0$ then the fields $\phi_1^\pm$ and $\chi_1$ become Goldstone bosons and the other fields become physical. With $\tau_2=0$, the Yukawa interactions are described by the interaction of $\phi_1$ with fermions (as $\phi_2$ does not couple to fermions but appears in loops). 
The Higgs and Yukawa Lagrangian in this case violate the $Z_2$ symmetry given by eq. (\ref{1stZ2symmetry}) but respect that given by eq. (\ref{2ndZ2symmetry}) only when $\tau_2=0$. Thus, parity in $\phi_2$ is conserved. This makes the lightest field of $\phi_2$ stable. 
Without this stability, we would have a model for accelerated expanding Universe for a limited time depending upon the decay width (life time) of the dark energy field(s), because if field(s) can decay then it ceases to provide accelerated expansion after its decay to the other fields. It would then require fine tuning to make it stable enough till the current age of the Universe and then to ``switch off''.
The masses of the fields in this vacuum are
\begin{equation}\begin{array}{rcl}\label{inertvacuumHiggsmasses}
		&&
		m_{\eta_1}^2=\rho_1 \Lambda_1 m_{11}^2~,
		\\ &&
		m_{\eta_2}^2=\rho_2 \Lambda_2 m_{22}^2 + \dfrac{\nu^2}{2}\left( \lambda_3+\lambda_4+\lambda_5 \right)	 ~,
		\\ &&
		m_{\chi_2}^2=\rho_2 \Lambda_2 m_{22}^2 + \dfrac{\nu^2}{2}\left( \lambda_3+\lambda_4-\lambda_5 \right)~,
		\\ &&
		m_{\phi_2^{\pm}}^2=2\rho_2 \Lambda_2 m_{22}^2 + \lambda_3 \nu^2~.
	\end{array}
\end{equation}


\section{Higgs fields as dark energy}\label{sec:3}
The Universe is homogeneous and isotropic at the cosmological scale, and its dynamics is described by the Friedmann equations given by eq. (\ref{1stFriedmannequation}) and eq.  (\ref{2ndFriemannequation}). Equation (\ref{2ndFriemannequation}) says that accelerated expansion will occur when $\omega_{eff}<-\frac{1}{3}$ where $\omega_{eff}=\Omega_{DE}\omega_{DE}+\Omega_{R}\omega_{R}+\Omega_{M}\omega_{M}$. For the field $\phi_2$ to be the dark energy field, it must bring $\omega_{eff}<-\frac{1}{3}$ within the history of the Universe (in fact just now $Z\approx 0.32$ when $\Omega_{DE}=0.7$ and $\Omega_{NR}=0.3$)
. For this purpose we need to solve the Euler-Lagrange equations, 
which are
\begin{eqnarray}\label{EulerLagrangeEquations}
	\partial_\mu
	\left( \dfrac{\partial(\sqrt{-g}{\mathscr{L}_{Higgs}})}{\partial(\partial_\mu \psi_i)} \right)-\dfrac{\partial(\sqrt{-g}{\mathscr{L}_{Higgs}})}{\partial \psi_i}=0,
\end{eqnarray}
where $\psi_i$ are different fields of doublets $\phi_1$ and $\phi_2$.

The Euler-Lagrange equations of motion in FRW Universe $(\sqrt{-g}=a(t)^3)$ for the fields $\phi_{2}^\pm$, $\eta_2$ and $\chi_2$ in this model are
\begin{equation}\begin{array}{rcl}\label{etaequationofmotion}
		&& \hspace{0cm} \ddot{\eta}_2 + 3\dfrac{\dot{a}}{a}\dot{\eta}_2 + \dfrac{1}{2}\eta_2\left(\nu^2\left(\lambda_3+\lambda_4+\lambda_5\right)+2 m_{22}^2\Lambda_2 \rho_2e^{m_{22}^2 \Lambda_2\left(\chi_2{}^2+\eta_2{}^2+2\phi_2^c{}^2\right) /2} \right. \\ && \hspace{2cm} \left. +\lambda_2 \Lambda_4 \rho_4 e^{\lambda_2 \Lambda_4\left(\chi_2{}^2+\eta_2{}^2+2 \phi_2^c{}^2\right){}^2 /8} \left(\chi_2{}^2 + 2\phi_2^c{}^2\right)\right) \\
		&& \hspace{2cm}+ \dfrac{1}{2}\lambda_2\Lambda_4\rho_4e^{\lambda_2\Lambda_4\left(\chi_2{}^2+\eta_2{}^2+2 \phi_2^c{}^2\right){}^2 /8} \eta_2{}^3=0,
	\end{array}
\end{equation}
\begin{equation}\begin{array}{rcl}\label{chiequationofmotion}
		&& \hspace{0cm}\ddot{\chi }_2+3 \dfrac{\dot{a}}{a}\dot{\chi }_2+\dfrac{1}{2}\chi _2 \left(\nu ^2\left(\lambda _3+\lambda _4-\lambda _5\right)+2 m_{22}^2
		\Lambda _2 \rho _2e^{m_{22}^2 \Lambda _2\left(\chi _2{}^2+\eta _2{}^2+2 \phi _2^c{}^2\right) /2} \right.\\ 
		&& \hspace{2cm} \left. +\lambda _2 \Lambda _4 \rho _4e^{\lambda _2 \Lambda _4\left(\chi _2{}^2+\eta _2{}^2+2 \phi _2^c{}^2\right){}^2 /8}\left(\eta _2{}^2+2\phi _2^c{}^2\right)\right) \\ &&
		\hspace{2cm}+\dfrac{1}{2}\lambda _2
		\Lambda _4 \rho _4e^{\lambda _2 \Lambda _4\left(\chi _2{}^2+\eta _2{}^2+2 \phi _2^c{}^2\right){}^2 /8}\chi _2{}^3=0,
	\end{array}
\end{equation}
\begin{equation}\begin{array}{rcl}\label{chargedHiggsequationofmotion}
		\ddot{\phi }_2^c+3 \dfrac{\dot{a}}{a}\dot{\phi }_2^c &+& \phi _2^c \left(\nu ^2 \lambda _3+2 m_{22}^2 \Lambda _2 \rho _2e^{m_{22}^2 \Lambda _2\left(\chi _2{}^2+\eta _2{}^2+2 \phi _2^c{}^2\right) /2} \right. \\ &+& \left. \lambda_2\Lambda_4\rho_4e^{\lambda_2\Lambda_4\left(\chi_2{}^2+\eta_2{}^2+2\phi_2^c{}^2\right){}^2 /8} \left(\chi_2{}^2+\eta_2{}^2+2\phi_2^c{}^2\right)\right)=0,
	\end{array}
\end{equation}		
where $c$ is $+$ or $-$.
The energy density and pressure after expansion of UDW-2HDM Higgs Lagrangian for physical fields become
\begin{equation}\begin{array}{rcl}\label{rhoPeta2}
		\rho _{\text{DE}}/\text{P}_{\text{DE}}= && \dfrac{1}{2}\dot{\chi }_2{}^2+\dfrac{1}{2} \dot{\eta }_2{}^2+\dfrac{1}{2}\dot{\phi }_2^c{}^2\pm \left(\rho _1+\rho _3 +\dfrac{1}{2} \nu ^2\lambda _3\phi _2^c{}^2\right.\\ && \left.+\dfrac{1}{4}\nu ^2\left(\lambda _3+\lambda _4+\lambda _5\right)\eta _2{}^2+\dfrac{1}{4}\nu ^2\left(\lambda _3+\lambda _4-\lambda _5\right)\chi _2{}^2 \right. 
		\\ && +\rho _2e^{m_{22}^2 \Lambda _2\left(\chi _2{}^2+\eta _2{}^2+2\phi _2^c{}^2\right) /2} \\ && \left.+\rho _4e^{\lambda _2 \Lambda _4\left(\chi _2{}^4+2 \chi _2{}^2 \eta _2{}^2 + \eta _2{}^4 +4\chi _2{}^2 \phi _2^c{}^2 +4 \eta _2{}^2 \phi _2^c{}^2+4\phi _2^c{}^4\right) /8}\right).
	\end{array}
\end{equation}

For the cosmological evolution of the fields $\eta_2$, $\chi_2$ and $\phi_2^c$, the equations of motion (given by eqs. (\ref{etaequationofmotion}, \ref{chiequationofmotion}, \ref{chargedHiggsequationofmotion})) are solved with the Friedmann equations numerically in the flat Universe ($\kappa=0$). The initial conditions used are ${\eta_2}_{ini}=M_P$, ${\chi_2}_{ini}=M_P$, ${\phi_2^c}_{ini}=0$, ${\dot{\eta}_2}_{ini}=0$, ${\dot{\chi}_2}_{ini}=0$ and ${\dot{\phi}_2^c}_{ini}=0$. The masses of the Higgs bosons in the analysis are taken to be 
$m_{\eta_2}=m_{\chi_2}=1.0247\times 10^{-59}\text{ GeV},$ 
to set the evolution of the energy densities as observed. 
Note that after imposing $Z_2$ symmetry there are five parameters which determine the masses of Higgs fields. With $m_{\eta_1}=m_{H_{SM}}=125.7$GeV we have only one equations to determine the parameters values \footnote{No constraint on parameters from tree level MSSM has been imposed}. The Higgs fields masses in this analysis were calculated by eq. (\ref{inertvacuumHiggsmasses}) by imposing $m_{\eta_1}=m_{H_{SM}}=125.7$GeV and some the arbitrary choice of parameters since there was only one equations to determine all unknown free parameters. One important thing in our model is that the mass of the charged Higgs becomes arbitrary and thus any value of the charged Higgs fields will suffice.

The solution of the eqs. (\ref{etaequationofmotion}, \ref{chiequationofmotion}, \ref{chargedHiggsequationofmotion}) along with Friedmann equations is shown below in the graphs.
\begin{figure}[h!]
	\hspace{-1.0cm}
	\centering
	\includegraphics[width=1\linewidth]{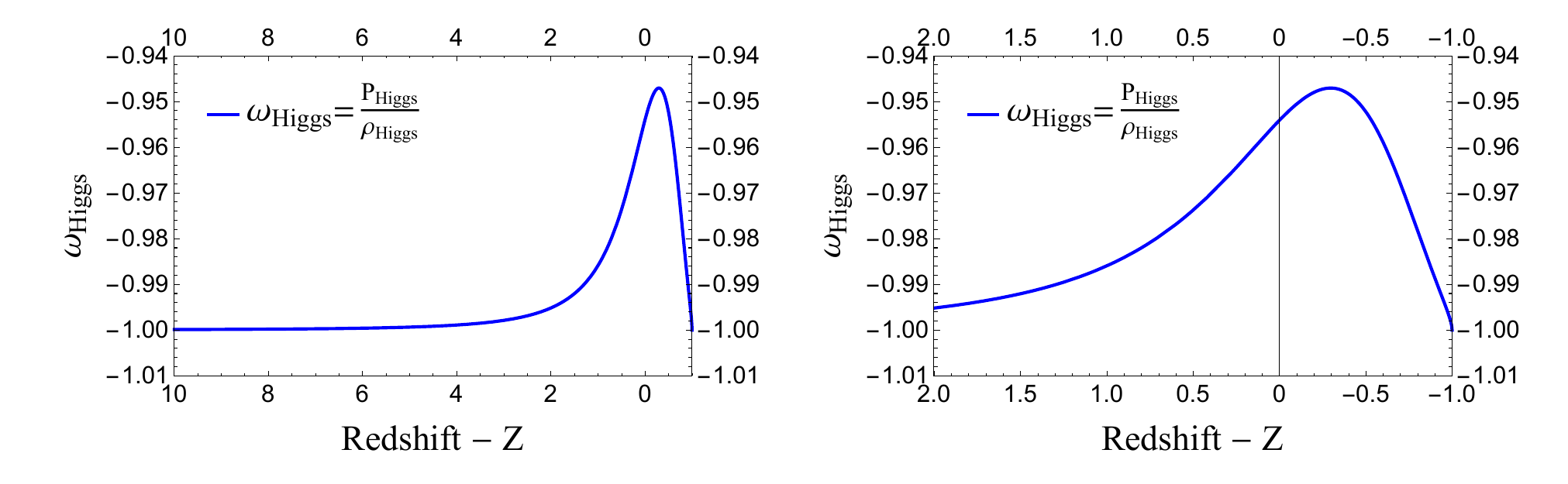}
	\caption{Effective equation of state parameter for Higgs fields $\omega_{Higgs}$, as seen it starts with $-1$ then evolves towards quintessence regime after large enough time it comes back at $-1$.}
	\label{fig:OmegaHiggs}
\end{figure}
During the evolution, $\omega_{Higgs}$ starts to increase and becomes $>-1$ as shown in fig. (\ref{fig:OmegaHiggs}). In the very late (future) Universe ($Z\ll0$), the fields come to rest at the minimum of the potential and a period with $\omega=-1$ is reachieved to give an accelerating Universe similar to a pure cosmological constant. Since $\omega_{Higgs}\ngtr-1/3$ at any time in the evolution, after $\omega_{eff}$ becomes $<-1/3$, we get an 
exponentially accelerating Universe.

As discussed before, the Higgs field stability is provided by imposing $Z_2$ symmetry. The lightest Higgs fields, $\eta_2$ and $\chi_2$, do not decay into any other Higgs field (since these fields are lighter than the SM-like and charged Higgs) or into fermions (since they do not couple to them at tree level).
\begin{figure}[h!]
	\hspace{-1.0cm}
	\centering
	\includegraphics[width=1\linewidth]{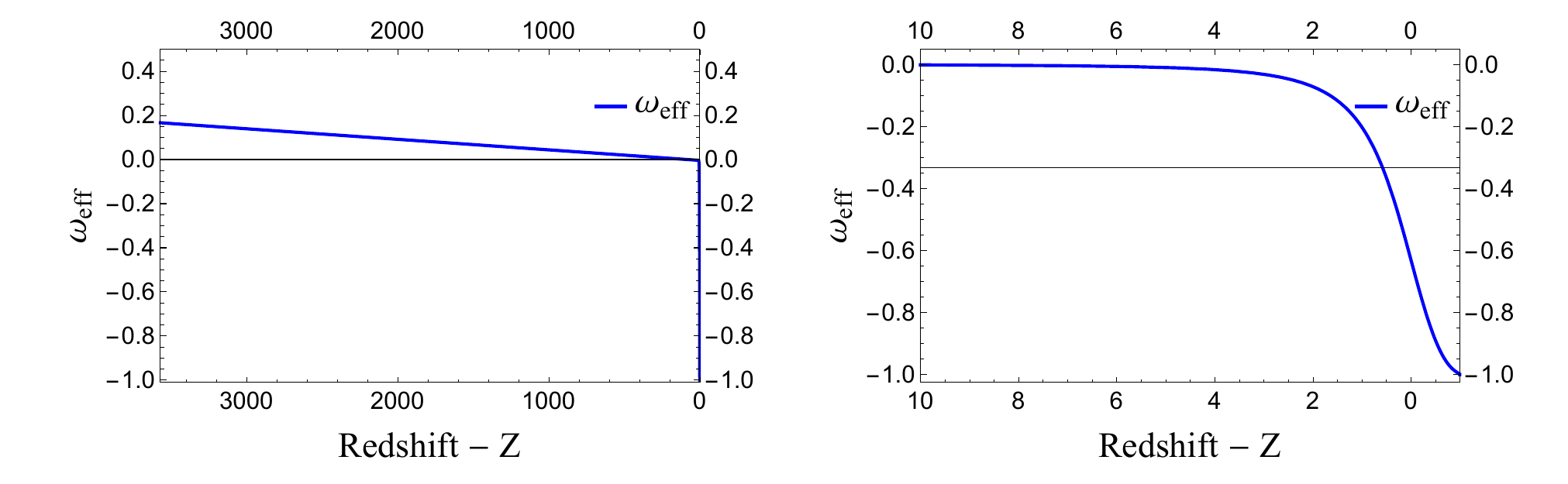}
	\caption{Effective equation of state parameter $\omega_{eff}=\Omega_{DE}\omega_{DE}+\Omega_{R}\omega_R+\Omega_{M}\omega_M$.
	}
	\label{fig:Omegaeffective}
\end{figure}	
\begin{figure}[h!]
	\centering
	\includegraphics[width=0.7\linewidth]{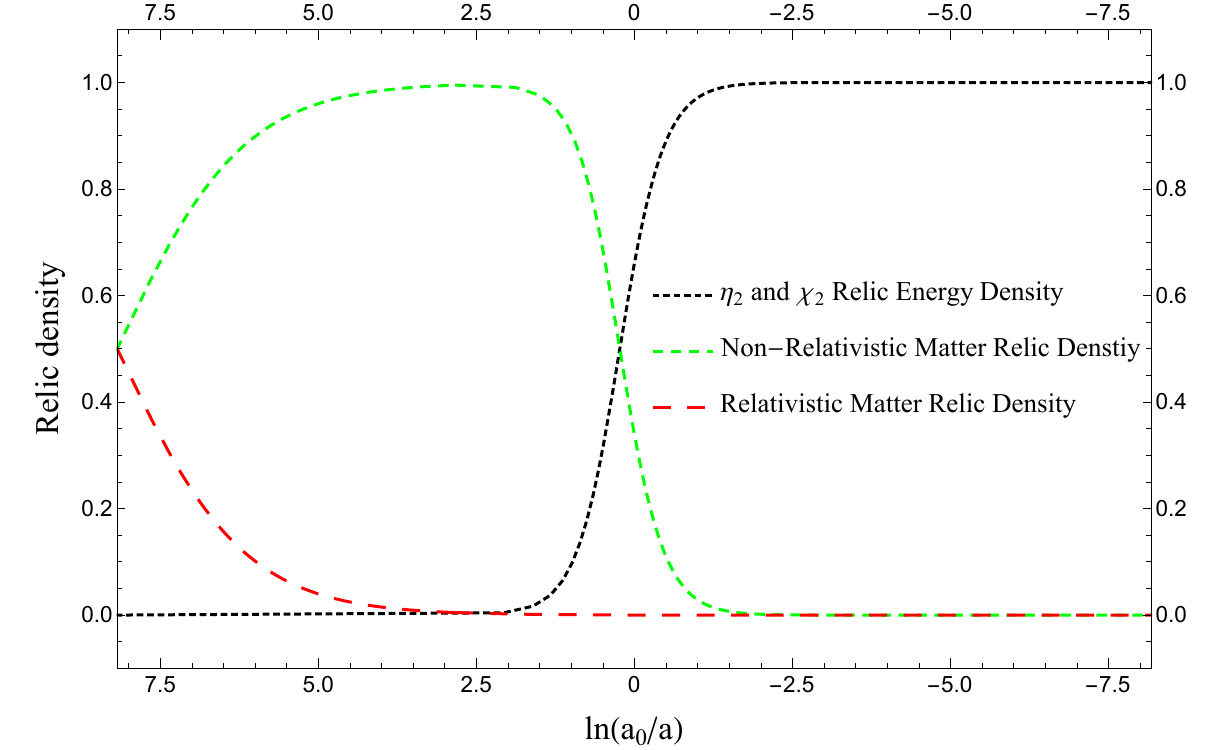}
	\caption{Relic densities of different components as a function of time.}
	\label{fig:relicdensities}
\end{figure}

The $\omega_{eff}$ in the fig. (\ref{fig:Omegaeffective}) starts from $\approx0.167$ (set by initial conditions $\Omega_{{Higgs}_{int}}=0$ and $\Omega_{{NR}_{int}}=\Omega_{R_{int}}=0.5$; NR: non-relativistic matter and R: relativistic matter) and decreases as the relativistic matter's energy density decreases (shown in fig. (\ref{fig:Omegaeffective}) for $\omega_{eff}$ and fig. (\ref{fig:relicdensities}) for relic densities). Initially, the relic density of non-relativistic matter increases while the relic density of relativistic matter decreases. The relic density of dark energy field approximately remains negligible in the initial stage of evolution. The time period when non-relativistic matter dominates with its relic density $\Omega_{NR}\approx 1$ is when the Universe decelerates at the highest rate 
as non-relativistic matter domination pulls 
things inwards more than the outwards Higgs negative pressure. When $\omega_{eff}=0$ the weighted negative pressure of dark energy fields and positive (inwards) pressure of non-relativistic matter cancel each other. After that $\omega_{eff}$ starts to decrease as the non-relativistic energy density decreases and the Higgs relic energy density increases (shown in fig. (\ref{fig:relicdensities})). From this time on the Higgs negative pressure dominates and $\omega_{eff}$ eventually settles down to $-1$.
%
Note that the initial conditions for the charged field took the dark energy (vacuum) not to be charged.
\footnote{As the charged field does not contribute to relic density, this does not mean that it can not exist, it can appear in the loop process.}
\section{Decay(s) of Higgs field(s)}\label{sec:4}
At the tree level the doublet $\phi_2$ in inert UDW-2HDM discussed above (whose component fields are the candidate for dark energy) is only coupled to the doublet $\phi_1$ which acts in an identical way as SM Higgs field.

The interaction Lagrangian of Higgs fields of doublet $\phi_2$ with Higgs fields of doublet $\phi_1$ and gauge bosons (extracted from eq. (\ref{LH})) is given by eq. (\ref{Lint}).

To suppress the interaction of Higgs fields, $\eta_2$, $\chi_2$ and $\phi_2^\pm$, with the gauge bosons the idea is that the SU(2) doublet $\phi_2$ is very weakly (different than $\phi_1$) coupled with the gauge bosons, thus we have $g_2\ll g_1$ and $g_2'\ll g_1'$.

The decay width of the Higgs to a pair of Higgs scalars, using only the on-shell width, is given by 
\cite{2hdmc,Eriksson2010985}
\begin{equation}\label{eta1eta2eta2}
\Gamma(H_i\longrightarrow H_j H_k) = (2-\delta_{jk})m_{H_i}\dfrac{|C_{H_i H_j H_k}|^2}{32\pi}\sqrt{f\left(1,\dfrac{m_{H_j}^2}{m_{H_i}^2},\dfrac{m_{H_k}^2}{m_{H_i}^2}\right)}~~,
\end{equation}
where
\begin{equation*}\label{function}
f\left(1,\dfrac{m_{H_j}^2}{m_{H_i}^2},\dfrac{m_{H_k}^2}{m_{H_i}^2}\right)=\left(1-\dfrac{m_{H_j}^2}{m_{H_i}^2}-\dfrac{m_{H_k}^2}{m_{H_i}^2}\right)^2-4~\dfrac{m_{H_j}^2}{m_{H_i}^2}~\dfrac{m_{H_k}^2}{m_{H_i}^2}~,
\end{equation*}
and $C_{H_i H_j H_k}$ is the coupling of different Higgs bosons $H_i$, $H_j$ and $H_k$.

\begin{equation}
\begin{array}[b]{rcl}\label{Lint}
\mathscr{L}_I &=& \dfrac{\nu }{2} \eta _1\eta _2{}^2 \left(\lambda _3+\lambda _4+\lambda _5\right)+\dfrac{\nu }{2} \eta _1\chi _2{}^2 \left(\lambda _3+\lambda
_4-\lambda _5\right)+\nu  \lambda _3\eta _1 \phi _2^c{}^2 \\ && +\dfrac{1}{4}\eta _1{}^2\eta _2{}^2 \left(\lambda _3+\lambda _4+\lambda _5\right)+\dfrac{1}{4}\eta
_1{}^2\chi _2{}^2 \left(\lambda _3+\lambda _4-\lambda _5\right) \\ && +\dfrac{1}{2}\eta _1{}^2 \phi _2^c{}^2 \lambda _3+\dfrac{1}{4}\eta _2{}^2\chi _2{}^2\left(m_{22}^4
\Lambda _2^2 \rho _2+\lambda _2 \Lambda _4 \rho _4\right) \\ && +
\dfrac{1}{2}\eta _2{}^2\phi _2^c{}^2 \left(m_{22}^4 \Lambda _2^2 \rho _2+\lambda _2 \Lambda _4 \rho _4\right)+\dfrac{1}{2}\chi _2{}^2 \phi _2^c{}^2
\left(m_{22}^4 \Lambda _2^2 \rho _2+\lambda _2 \Lambda _4 \rho _4\right)
\vspace{0.2cm} \\ &&
+\dfrac{g_2^2g_2'{}^2}{(g_2'{}^2+g_2^2)}\phi_2^+\phi_2^-A_{\mu}^2 +\dfrac{(g_2'{}^2+g_2^2)}{8}\eta_2^2Z_{\mu}^2+\dfrac{(g_2'{}^2+g_2^2)}{8}\chi_2^2Z_{\mu}^2
\vspace{0.2cm} \\ &&
+\dfrac{(g_2^2-g_2'{}^2){}^2}{4(g_2'{}^2+g_2^2)}\phi_2^+\phi_2^-Z_{\mu }^2+\dfrac{g_2g_2'(g_2^2-g_2'{}^2)}{(g_2'{}^2+g_2^2)}\phi_2^+\phi_2^-A_{\mu}Z_{\mu}
\vspace{0.2cm} \\ &&
+\dfrac{g_2^2}{4}W^-{}_\mu W^+{}_\mu(\eta_2^2+\chi_2^2+2\phi_2^+\phi_2^-) \\ && +\dfrac{g_2{}^2g_2'}{2\sqrt{(g_2'{}^2+g_2^2)}}\eta_2A_{\mu}(\phi_2^+W^-{}_\mu+\phi_2^-W^+{}_\mu)
\vspace{0.2cm} \\ &&
+i\dfrac{g_2{}^2g_2'}{2\sqrt{(g_2'{}^2+g_2^2)}}\chi_2A_{\mu}(\phi_2^-W^+{}_\mu-\phi_2^+W^-{}_\mu) \\ && -\dfrac{g_2g_2'{}^2}{2\sqrt{(g_2'{}^2+g_2^2)}}\eta_2Z_\mu(\phi_2^+W^-{}_\mu+\phi_2^-W^+{}_\mu) \\ && -i\dfrac{g_2g_2'{}^2}{2\sqrt{(g_2'{}^2+g_2^2)}}\chi_2Z_\mu(\phi_2^-W^+{}_\mu-\phi_2^+W^-{}_\mu).
\end{array}
\end{equation}

From eq. (\ref{Lint}), $C_{\eta_1 \chi_2 \chi_2}=\dfrac{1}{2}\nu(\lambda_3+\lambda_4-\lambda_5)$, $C_{\eta_1 \eta_2 \eta_2}=\dfrac{1}{2}\nu(\lambda_3+\lambda_4+\lambda_5)=$ and $C_{\eta_1 \phi_2^+ \phi_2^-}=\lambda_3\nu$. 
The decay rates of $\eta_1$ to $\eta_2\eta_2$, $\chi_2\chi_2$, $\phi_2^-\phi_2^+$ for the masses used in the cosmological evolution determination are zero.


It should also be mentioned that the total decay width of the SM-like Higgs boson here is well within the bounds of mass resolution $\approx12\times10^{-3}$GeV of 
LHC \cite{PDG} 
for the values of the masses used in the cosmological evolution of the Higgs fields. One should also mention that the SM prediction of total decay width for the Higgs boson is $4.21\times10^{-3}$GeV with mass $126$GeV \cite{LHC-CSWG} and is $4.07\times10^{-3}$GeV with mass $125$GeV \cite{PDG}. In this model, we get three more decay channels of the SM-like Higgs, to the other Higgs bosons pair.

\section{Conclusion}\label{sec:Conclusion}
Scalar fields are among the possible candidates for the observed accelerated expansion of the Universe. In this article, we have argued that the Particle Physics developed so far must have something in, or minimally beyond, the SM which will explain the observed accelerated expansion of the Universe and hence will serve as a dark energy candidate. 
Here we assumed that dark energy is actually some scalar field which is present as the Higgs in a model where the potential has the non-degenerate vacua, we called this model uplifted double well two-Higgs doublet model (UDW-2HDM).

We found that if the present Universe is described by the true vacua of UDW-2HDM then the component fields of the second doublet $\phi_2$ (which acts as the inert doublet) can be one possible candidate for the dark energy. As the present contribution of the dark energy to the critical energy density is about $0.7$, this value is obtained by taking the mass of the CP-even field's mass small ($O(10^{-59})$GeV). The most important thing is that with the initial conditions set, the mass of the charged ($\phi_2^\pm$) field becomes arbitrary. Hence this model will fit for any value of mass of $\phi_2^\pm$. One also needs to keep in mind that the values of masses were chosen arbitrarily so as to get dark energy relic density $\approx 0.7$. Changing the values of the masses, the relic density does not change much. 

It should also be mentioned that if we remove the $Z_2$ symmetry, the second Higgs doublet does not remain inert. Thus in the case of $Z_2$ violation (soft or hard), the CP-even Higgs fields will mix by an angle $\beta$. In that case a new parameter ($\beta$) will arise in the theory. Obtaining a dark energy candidate in that model will require fine tuning in the Yukawa interactions in such a way that either the dark energy field does not couple or couple very weakly with the fermions.

\section{Acknowledgment}
This work is supported by 
\emph{Higher education commission (HEC) of Pakistan} under the project no. NRPU-3053.

\bibliographystyle{ws-ijmpd}
\bibliography{Muhammad-Usman.bib}


\end{document}